\documentclass[useAMS,usenatbib]{mn2e}

\usepackage{graphicx,times}

\def\deg{$^{\circ}$}
\def\solm{M$_{\odot}$}
\def\kms{km s$^{-1}$}
\def\kmskpc{km s$^{-1}$ kpc$^{-1}$}

\title[Regular motions in double bars. II. Survey of 23 models]
{Regular motions in double bars. II. Survey of trajectories and 23 models}

\author[Witold Maciejewski and E. Athanassoula]
{Witold Maciejewski$^{1}$\thanks{E-mail:wxm@astro.livjm.ac.uk} and 
E. Athanassoula$^{2}$\\
$^{1}$Astrophysics Research Institute, Liverpool John Moores University, Twelve
Quays House, Egerton Wharf, Birkenhead, CH41 1LD\\
$^{2}$Laboratoire d'Astrophysique de Marseille (LAM), UMR6110, CNRS/Universit\'e
de Provence, 2 Place Le Verrier, 13248 Marseille C\'edex 04, France}

\begin{document}

\maketitle

\begin{abstract}
We show that stable double-frequency orbits form the backbone of double bars,
because they trap around themselves regular orbits, as stable closed periodic 
orbits do in single bars, and in both cases the trapped orbits occupy similar 
volume of phase-space. We perform a global search for such stable 
double-frequency orbits in a model of double bars by constructing
maps of trajectories with initial conditions well sampled over the available
phase-space. We use the width of a ring sufficient to enclose a given map as
the indicator of how tightly the trajectory is trapped around a 
double-frequency orbit. We construct histograms of these ring widths in order 
to determine the fraction of phase-space occupied by ordered motions. 
We build 22 further models of double bars, and we construct histograms
showing the fraction of the phase-space occupied by regular orbits
in each model. Our models indicate that resonant coupling between the bars 
may not be the dominant factor reducing chaos in the system.
\end{abstract}


\section{Introduction}
In this series of papers, we study the dynamics of galaxies with double bars, 
i.e. systems, where a small bar is nested inside a large-scale outer bar, and 
the two bars rotate with two different pattern speeds. Such double bars belong 
to a general class of oscillating potentials, with the oscillation period being
the time between two consecutive alignments of the bars. An oscillating 
system in $N$ dimensions is equivalent to an autonomous system in $N+1$ 
dimensions (e.g. Lichtenbeg \& Lieberman 1992, Louis \& Gerhard 1988). 
In the first paper of the series (Maciejewski \& Athanassoula 
2007, hereafter Paper I) we showed that  
double bars have no continuous families of
closed periodic (single-frequency) orbits; instead, their fundamental
kind of orbits have two frequencies. 
These double-frequency orbits constitute a subset of the regular orbits in 
the plane of a doubly barred galaxy and, if stable, they are 
 surrounded by  
regular (three-frequency) orbits in the same way as stable closed periodic 
orbits in a single bar are surrounded by two-frequency orbits.
In both cases, 
the trapped regular orbits oscillate around the parent orbit. In Paper I, we 
showed an example of the parent stable double-frequency orbit in double bars 
(fig.2 in Paper I), and an example of a regular orbit that is trapped around 
it (the right panels of fig.1 in Paper I).

Trajectories in a pulsating potential are difficult to study, because they
do not close in any reference frame. Some information about their eccentricity
and their alignment with the bars can be obtained from measuring their maximum 
extent in the direction of bar's major and minor axes in the frame of the bar
(El-Zant \& Shlosman 2003). Maciejewski \& Sparke (2000, hereafter MS00) 
proposed another way to study such trajectories. They used maps, each built
out of a discrete set of points on a given trajectory separated by the interval
equal to the oscillation period. In double bars, maps of trajectories are 
generated by writing the position on the particle's trajectory every time 
interval equal to the relative period of the bars. In Paper I, we showed that
maps of double-frequency orbits created by this construction form closed 
curves, which we call loops, following Maciejewski \& Sparke (1997), 
who originally defined and named these curves. Loops in double bars
can be studied in the  
same way as closed periodic orbits in the frame of a single bar. Regular 
orbits trapped around double-frequency orbits (referred to in this paper as 
trapped trajectories) map onto rings enclosing the loops. One can follow the 
transformation of a loop or a ring as bars rotate through one another to see 
whether it supports either bar, and we follow this approach here to 
study trajectories in the pulsating potential of double bars. Some trajectories
in double bars can be trapped by one bar for a limited period of time, and
then move to chaotic region of phase space before being trapped by another bar.
In our approach, such trajectories will not show as trapped by any bar, and
therefore we may underestimate the orbital support of double bars.

As in Paper I, we limit ourselves to studying trajectories 
in the plane of the galaxy, since we extensively use the concept of the ring 
in the maps of trajectories, which is well defined only in the 
plane of the galaxy. Consequently, we consider only trajectories
with no initial velocity component in the direction perpendicular 
to the galactic plane. Stable double-frequency orbits in double bars 
can be found by searching for trajectories whose maps can be enclosed 
by rings of the smallest possible width. As we showed in Paper I, the 
smaller the ring width in the map of a trajectory, the less oscillations in
this trajectory around the parent orbit. Here, we explore a multi-dimensional
phase-space of initial conditions, in which families of double-frequency orbits
can be localized by one-dimensional stretches of near-zero ring widths. We 
start the search for families of double-frequency orbits in double bars from 
Model 2 of MS00, which we will call the Reference Model in this paper. We 
chose this starting point because the Reference Model 
is so far the best understood dynamically plausible model of double bars, i.e. 
the loops there follow the motion of the outer and the inner bar throughout the
whole extent of the bars. In this model, we explore thoroughly the phase-space 
of initial conditions, achieving completeness similar to that of surfaces of 
section in a single bar, in order to find all orbital families that may play a 
role in supporting double bars. We also construct other models of double 
bars, for which we explore the extent of only major families of 
double-frequency orbits.
In each model, we estimate the fraction of the phase-space occupied by regular 
motions. In double bars, a piling up of resonances created by each bar
is likely to occur, which leads to considerable chaotic zones. In order to 
sustain the two bars dynamically, a resonant coupling may occur, as that 
proposed by Tagger et al.~(1987) and Sygnet et al.~(1988), so that a resonance 
generated by 
one bar overlaps with another caused by the other bar. Here we 
estimate the extent of chaos both in resonantly coupled and uncoupled double 
bars.

In Section 2, we compare the distribution of ring-widths in maps of 
trajectories in a single bar and in double bars using the Reference Model.
In order to determine the fraction of the phase-space occupied by ordered 
motions, we construct histograms of the ring width. In Section 3, 
we analyze maps at various relative positions of the bars, and we search for 
asymmetric loops by considering initial velocities with a radial component.
In Section 4, we build 
models whose characteristics depart from those of the Reference Model, and 
study regular motions in these models. By comparing ring-width diagrams and
histograms we select models of double bars that are most plausible 
dynamically. In Section 5, we discuss the orbital response to varying 
parameters of the models, and we compare the predictions of our models
with N-body simulations.

\section{Comparison of regular motions in single and double bars}
In a single bar, there are four initial conditions for a trajectory in the 
galaxy plane: two for the initial position and two for the initial velocity.
Since one integral of motion, the Jacobi integral, is conserved in a rotating 
bar, three initial conditions are needed for a trajectory with a given Jacobi
integral. If this trajectory covers the whole 2$\pi$ angle in azimuth, no 
generality is lost if its origin is assumed to be at one given azimuthal angle,
e.g. on the minor axis of the bar. This reduces the number of initial 
conditions to two, and allows the study of motions in a single bar by 
constructing surfaces of section. 

In a double bar, there is one more initial condition for a trajectory in the
galactic plane, the initial relative angle of the bars, which brings
the total number of initial conditions to five. Moreover, the Jacobi integral 
is no longer conserved, hence surfaces of section cannot be constructed.
Like in a single bar, the number of initial conditions can be reduced by one
after assuming the initial azimuthal angle of the trajectory. As a first step,
we start the trajectory on the minor axis of the aligned bars, hence we
fix one more initial condition: the relative angle of the bars. Later on, we
will release this assumption. 
We start with the bars aligned, because only for bars aligned (parallel)
or perpendicular, the potential preserves the symmetry with respect to the 
bars' axes, present in the case of a single bar. Nevertheless, this still 
leaves us with three independent initial conditions, instead of two in a single
bar, and we have to explore the phase-space of initial conditions differently. 

We start our search of fundamental orbits in double bars with exploration of 
trajectories that have no initial radial velocity component, because in a 
single bar the fundamental x1 and x2 orbital families (we use the generally 
accepted notation from Contopoulos \& Papayannopoulos 1980), and many other 
important families there are symmetric with respect to the axes of the bar and
have no radial velocity component when crossing the minor axis. With this 
constraint, we expect to recover double-frequency orbits, whose maps are 
symmetric around the axes of the bars when the two bars are parallel. An 
additional benefit of imposing this constraint is that we are left with only 
two free parameters to vary for the initial conditions: the starting location 
of the particle on the minor axis of the aligned bars, and the value of its 
initial tangential velocity.

\begin{figure*}
\centering
\includegraphics[width=0.97\linewidth]{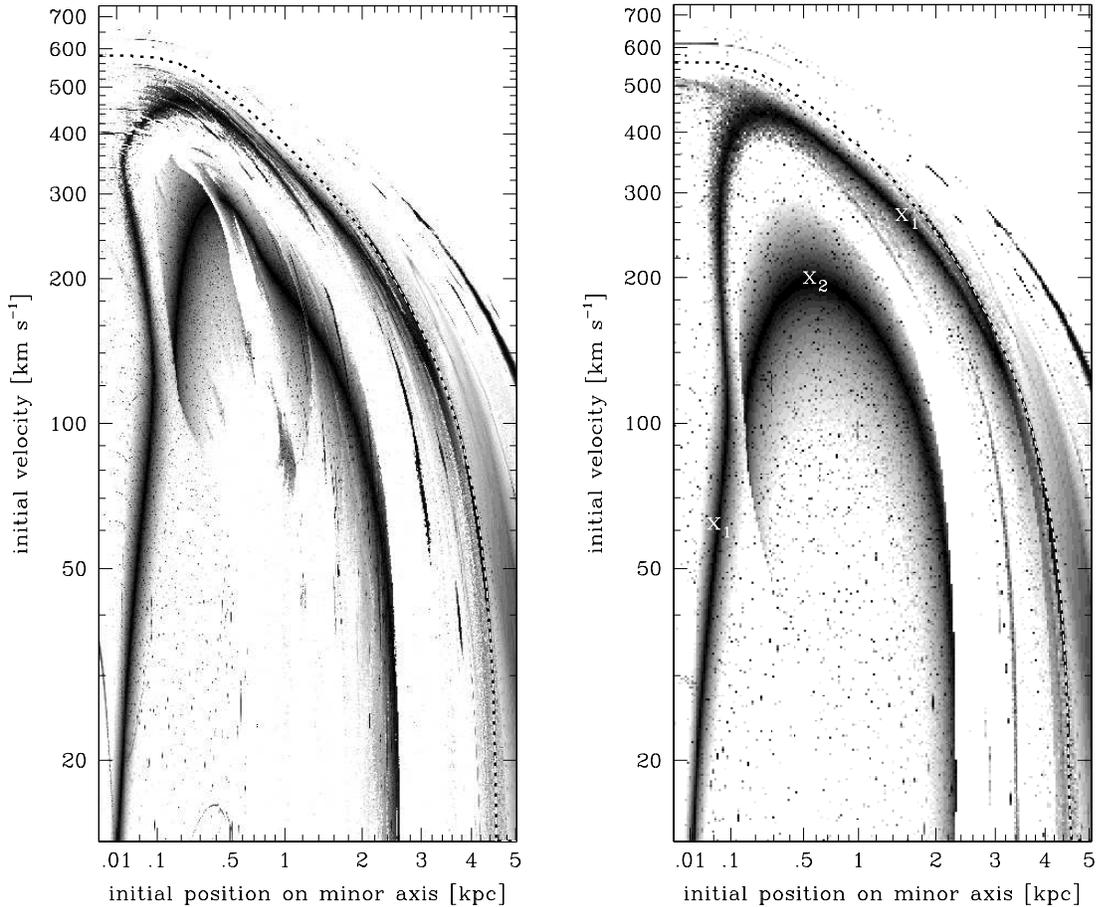}
\vspace{-119mm}
\caption[]{{\bf Left-hand panel:} Width of the ring enclosing maps 
of trajectories in the Reference Model, plotted in greyscale as a function 
of the particle's initial position on the minor axis of the aligned bars, 
and of the value of its initial velocity, perpendicular to this axis. Darker 
color means smaller width, and widths above 50 per cent of the ring's
average radius are shown in white. The white line with black dots marks values 
of initial conditions above which energetics
allows the particle to pass the L1 Lagrange point of the outer bar (see text). 
{\bf Right-hand panel:} Same as in the left-hand panel, but for the 
trajectories in a single bar generated from the Reference Model 
by incorporating the mass of the inner bar into the spheroid. Two dark 
stripes corresponding to the trajectories trapped around the x1 and x2 
periodic orbits are marked.}
\label{f1}
\end{figure*}

\subsection{The double-bar Reference Model and the single-bar model derived 
from it}
Our Reference Model (Model 2 in MS00), consists of a spheroid, a disc and two
bars, with, respectively, a Hubble, a Kuzmin-Toomre and a Ferrers profile, as 
in the single bar models of Athanassoula (1992a).
The exponent in the Ferrers formula is 2. The L1
Lagrangian point for the outer, main bar is located 6 kpc from the galaxy
centre. It roughly corresponds to the corotation radius of the outer bar.
The semi-major axis of the main bar is also 6 kpc, hence the bar is rapidly
rotating. The semi-major axis of the inner, secondary bar is 1.2 kpc, hence
its linear size is 5 times smaller  than that of the outer bar. The axial 
ratios are 2.5 for the outer bar and 2 for the inner bar. The quadrupole moment
of the outer bar is $4.5 \times 10^{10}$ \solm kpc$^2$, with the maximum ratio 
of the tangential to the radial force from the total mass distribution being 
about 20 per cent, which indicates a medium-strength bar. The mass of the inner
bar is 15 per cent 
of that of the outer one. Pattern speeds of the bars are not commensurate 
(36.0 \kmskpc\ for the big bar, and 110.0 \kmskpc\ for the small one). This
last value places the corotation of the inner bar at 2.2 kpc, at the location
of the Inner Lindblad Resonance of the outer bar, hence a resonant coupling
exists between the bars in this model. On the other hand, the corotation of 
the inner bar is far beyond its end, hence the inner bar is not rotating 
rapidly.

Mass distribution in our models is constructed by assuming the parameters of 
the disk and the spheroid, and then extracting from the spheroid the mass that 
is needed to create bars of the required size, quadrupole moment and density
profile. In this way, the total mass of the system remains constant, and all
models have the same total mass. For the
single-bar model derived from the Reference Model, the mass of the secondary 
bar is not extracted, and it remains in the spheroid.

\subsection{The ring-width diagrams}
As already mentioned 
at the start of Section 2, we first reduce the number of initial conditions of 
trajectories that we want to explore to two: the initial position along the 
minor axis of the aligned bars, and the initial velocity perpendicular to 
this axis. Now, we can construct two-dimensional diagrams of any variable 
that characterizes the resulting trajectory, as a function of the values of
initial conditions plotted on the two axes of the diagram. The variable that
allows finding double-frequency orbits and trajectories trapped around them
in double bars is the width of the ring that encloses the map of the
trajectory. The minimal width indicates a double-frequency orbit,
and small widths for adjacent initial conditions indicate trajectories 
trapped around the double-frequency orbit. 

We construct the maps from trajectories followed for 399 alignments of the 
bars, therefore each map consists of 400 points. For each map we measure the 
width of the ring within which these points fall. The method of calculating
the ring width, based on the method proposed by MS00,
is explained in detail in Paper I. To summarize it, the ring 
width is defined as the median radial spread of points among a number (usually 
40) of equal azimuthal sectors covering the whole 2$\pi$ angle, normalized by 
the average radial coordinate of the points that constitute the map. When this
normalized ring width is much smaller than 1, one may expect a trapped 
trajectory, while there is no sign of the trajectory being trapped when it is 
close to 1. Moreover, in Paper I, we showed that the numerical value of the 
width of the ring indicates how tightly the particle is trapped around its 
parent double-frequency orbit. 

In the left panel of Fig.1, we show the diagram of the ring width for the 
Reference Model. We sample the space of initial conditions for 400 starting 
positions and 800 starting velocity values, hence 320,000 ring widths are 
displayed in greyscale. The ring width is measured for the maps obtained at 
the moments when the major axes of the bars overlap.
Darker shading corresponds to smaller widths, while widths higher than 
0.5 are left white (trajectories that map onto rings of widths higher
than 0.5 are
unlikely to be trapped around regular orbits -- see fig.5 in Paper I).
This plot is essentially fig.8a from MS00, albeit at a much 
higher resolution. Dark shadings form clear stripes, darkest in their centres.
Taking cuts through these stripes for a constant initial position on the 
minor axis will produce diagrams of ring width like the one from the lower 
panel of fig.5 in Paper I. In Paper I, we showed that trajectories entering 
that diagram are trapped around a double-frequency orbit, whose map has the 
smallest ring width. Therefore in the two-dimensional representation of Fig.1,
the darkest lines in the interior of the stripes are likely to correspond to
maps of double-frequency orbits, since the measured ring width of the loop is
very small. The remainder of each stripe that surrounds those lines indicates
regular trajectories trapped around those double-frequency orbits.
In Paper III (Maciejewski et al. 2008, in preparation), we will
prove that this is in fact the case.

The main families of loops in double bars are related to the main orbital
families in a single bar (MS00). In the right panel of Fig.1, we show the 
diagram of the ring width for the single-bar model derived from the Reference
Model, as described in Section 2.1. The mapping of trajectories is still
done every time period equal to the relative period of the bars in the 
Reference Model. We do this for consistency 
with the double bar work, although in a single bar the frequency of writing 
is irrelevant, since the loop is identical to the closed periodic orbit. 
Like in the case of double bars, we followed the particle for a time 
equal to 399 relative periods of the bars in the Reference Model, writing 
its position 400 times at equal time intervals.

MS00 showed to which orbits in a single bar various features in the
ring-width diagram correspond, and we will expand this analysis in
Paper III. Here we only summarize that
trajectories with initial conditions from the darkest 'spine' of the stripe 
forming the lower arch in the single-bar diagram in Fig.1 belong to the x2 
orbital 
family. The stripe that makes the upper arch represents trajectories trapped 
around the x1 orbital family and orbital families related to it. A stripe of 
small ring width, emerging from the right side of the box at a velocity about 
120 \kms corresponds to orbits beyond the corotation of the bar, belonging 
to the outer 2:1 orbital family (see e.g. fig.11 in Sellwood \& Wilkinson 
1993). There are also other, secondary features in the ring-width diagram, like
grey stripes that do not include ring widths near zero, but their importance
for the dynamics of the system is small.

When comparing the two diagrams in Fig.1, one can notice that the two dark 
arcs, marking the x1 and x2 orbital families in the single bar, occur at 
similar locations in the diagram for the double bar, and they have similar 
appearance there. This confirms the finding of 
MS00 that in double bars there are double-frequency orbits that correspond 
to closed periodic orbits in single bars. Thus, one can use the diagram from 
the left panel of Fig.1 to single out families of double-frequency orbits. 
This will be done in Paper III. Major double-frequency orbits map onto loops,
whose appearance was given in MS00.

Despite general similarities, there are also significant differences 
between the diagrams in Fig.1. The diagram for the double bar looks less
regular, with a number of white stripes superimposed on the two dark arches
representing the two major orbital families. The most spectacular white stripe 
cuts through the top of the lower arch. White color means normalized ring 
widths above 50 per cent, hence most likely trajectories not trapped around any
double-frequency orbits. If a white stripe cuts through the dark arch, it
means discontinuity in stable double-frequency orbits there.

The other difference is that the dark arches look wider in the diagram for the 
single bar. This indicates that trajectories trapped around double-frequency 
orbits in a double bar occupy smaller volume of phase-space than trajectories 
trapped around closed periodic 
orbits in its corresponding single bar. On the other hand, the dark 
zones in the double bar still cover a large fraction of the diagram, hence
regular orbits trapped around double-frequency orbits occupy a considerable
fraction of phase-space there. 

\begin{figure}
\centering
\rotatebox{-90}{\includegraphics[width=.7\linewidth]{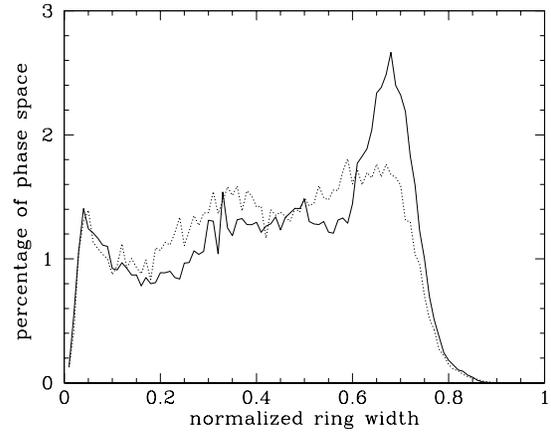}}
\caption[]{Histogram showing the percentage of phase-space occupied by
the trajectories that map onto rings of normalized widths indicated on the 
horizontal axis of the diagram, for the Reference Model (solid line), and 
the single-bar model (dotted line).}
\label{f2}
\end{figure}

\subsection{Histograms of ring width: phase-space volume occupied by regular motions}
Double-frequency orbits can form the backbone of double bars, if trajectories 
trapped around them occupy a significant amount of phase-space in the system. 
Here we compare how much of phase-space is occupied by trajectories trapped
around closed periodic orbits in a single bar, and those trapped around 
double-frequency orbits in double bars.

In order to quantify the amount of phase-space trapped around parent orbits 
in single and double bars, we created histograms of ring width. We considered
only trajectories confined within the corotation of the outer bar. It is well
known that in the case of a single bar, the Jacobi integral $E_J$ is constant,
and a particle is confined to remain inside the corotation of the bar if its
Jacobi integral is smaller than
the effective potential $\Phi_{\rm eff}$ at the Lagrange point $L_1$, 
$\Phi_{\rm eff}(L_1)$. Since for a particle with the position ${\bf r}$ and 
the velocity $\dot{\bf r}$
\begin{equation}
E_J({\bf r},\dot{\bf r}) = \frac{1}{2} |\dot{\bf r}|^2 + \Phi_{\rm eff}({\bf r}),
\end{equation}
the particle starting at the position ${\bf r}$ will remain inside the bar's 
corotation if the starting velocity is smaller than
\begin{equation}
v_{\rm {max}} = \sqrt{2(\Phi_{\rm eff}(L_1) - \Phi_{\rm eff}({\bf r}))}.
\end{equation}
Although the Jacobi integral is not conserved in double bars, the procedure 
above can serve as a good approximation there, because direct integration 
of trajectories shows that particles with initial velocity smaller
than $v_{\rm {max}}$ are confined within the corotation of the outer bar
also in double bars. This is because the inner bar has little impact on
the Jacobi integral of particles far outside of it, while particles 
at small radii, where the presence of the inner bar makes the Jacobi 
integral truly time-dependent, are already well confined within the corotation 
of the outer bar.

For both a single bar, and a double bar with the bars aligned (the starting
configuration), we calculated $v_{\rm {max}}$ as a function of the initial 
position on the minor axis of the bars. It is plotted in Fig.1 as the white
line with black dots superimposed, so that it can be seen against varying
background. It 
shows no significant difference between the case of one and two bars, and the 
results below remain unchanged regardless of whether we use the $v_{\rm {max}}$
derived for a single bar, or for two aligned bars, when analyzing the double
bar case.

\begin{figure*}
\centering
\includegraphics[width=\linewidth]{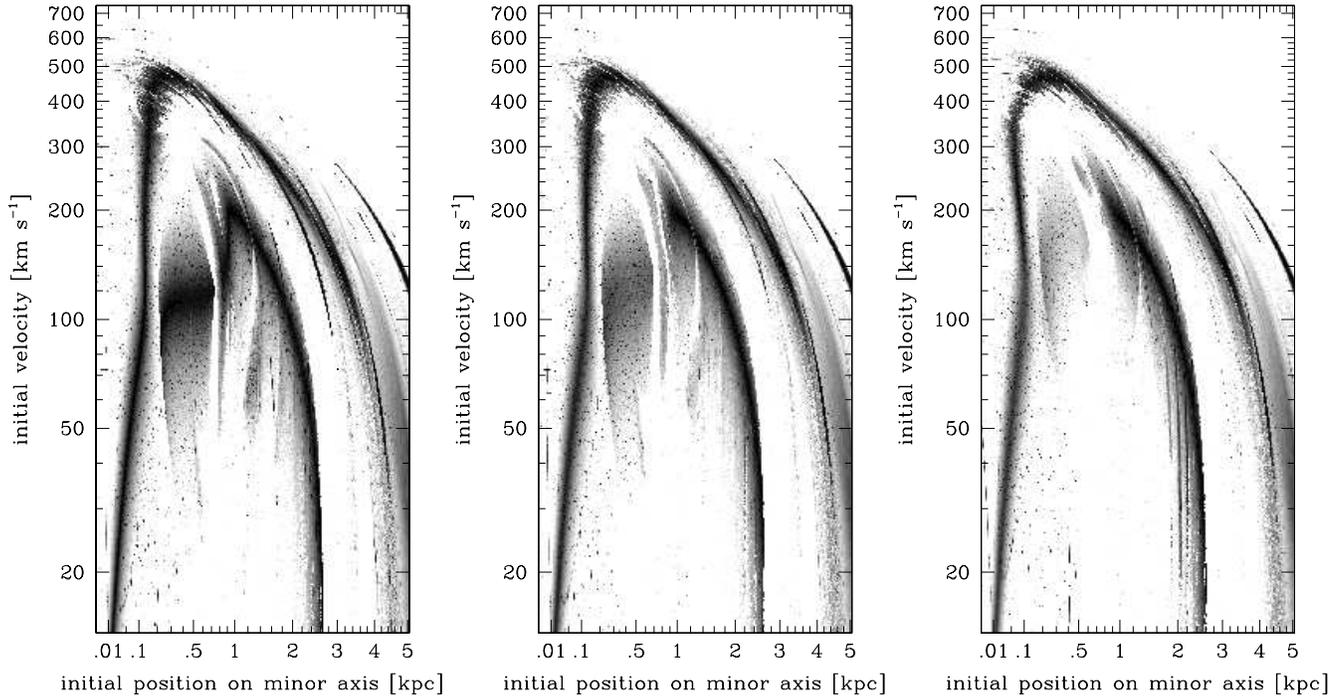}
\vspace{-161mm}
\caption[]{Ring-width diagrams in the Reference Model, as in the left-hand 
panel of Fig.1, but for the
trajectories starting at the moments when the angle between the two bars is 
$\alpha=0.5\pi$ ({\bf left-hand panel}), 0.4$\pi$ ({\bf central panel}) and 
0.2$\pi$ ({\bf right-hand panel}). The diagrams are plotted 
as a function of the particle's initial position along the minor axis of the 
outer bar, and of its initial value of velocity perpendicular to that axis.}
\label{f3}
\end{figure*}

As labels on the axes of the diagrams in Fig.1 indicate, we do not sample the
phase-space of initial conditions uniformly, hence the volume of the
phase-space $\Delta ({\bf r},\dot{\bf r})$ corresponding to a given initial 
condition is
\begin{equation}
\Delta ({\bf r},\dot{\bf r}) = (\Delta y/r_L) * (\Delta v/v_{\rm max}),
\end{equation}
where $\Delta y$ is the local spacing in the sampling of the initial position
on the minor axis, and $\Delta v$ is the local spacing in the sampling of the
initial velocity. The volume $\Delta ({\bf r},\dot{\bf r})$ of the phase-space
element is normalized by the radius of the Lagrange point $L_1$ for the outer 
bar, $r_L$, and by $v_{\rm {max}}$. The histogram is constructed for 100 bins 
of the ring width, each covering equal fractional width (1 per cent) by adding 
the normalized volumes (3) grouped in bins determined by their ring width. 
Since
we limit our search to particles remaining within the corotation of the outer 
bar, out of the 320,000 initial conditions sampled, we add to the bins only 
those, where the initial velocity is smaller than $v_{\rm {max}}$ for the 
given initial position on the minor axis (i.e. points located below the white 
line with black dots in the diagrams in Fig.1). 

In Fig.2, the histogram for the single bar is drawn with the dotted line, and
the solid line marks the histogram for the double bar. Starting from the left
end, both histograms show a local peak at the normalized ring width of about 
5 per cent.
This peak corresponds to trajectories well trapped around the parent orbits --
closed periodic orbits in a single bar and double-frequency orbits in double 
bars. The volume of phase-space occupied by trajectories mapping onto rings of 
width below 10 per cent is roughly the same in both histograms. This means that
double-frequency orbits in double bars are equally good in trapping around
themselves regular trajectories as closed periodic orbits are in single bars.
But this is only true about closely trapped trajectories. The volume of 
phase-space occupied by trajectories that map onto rings of larger width, 
between 20 and 60 per cent, is  
systematically lower in double bars than it is in a single bar. This 
'missing volume' reappears in double bars at ring widths between 60 and 80 
per cent as a separate component of the histogram.
Its quasi-Gaussian shape indicates chaos, as we will confirm in Paper III.
The absence of such a component in
the histogram for a single bar indicates a much smaller fraction of chaotic 
motions there. A small fraction of chaotic orbits is indeed expected for a bar 
of relatively small axial ratio 2.5, and rather high Ferrers index of 2.

From the above analysis, we conclude that the amount of phase-space well
trapped around double-frequency orbits in the Reference Model of a double bar
and around closed periodic orbits in the corresponding model of a single bar
is comparable, hence double-frequency orbits can provide the backbone
for double bars in the same way as closed periodic orbits do in a single bar.
On the other hand, the overall fraction of phase-space occupied by chaotic 
orbits in the model with
double bars is much higher than in a single bar, which confirms the expectation
that double bars introduce chaos. Note, however, that in our model, despite of
large fractions of phase-space being chaotic, the regions of phase-space, 
where orbits forming the backbone of the system reside, remain largely regular.

\section{Search for other regular orbits in the Reference Model}
So far we considered only trajectories starting on the minor axis of the 
aligned bars, with the initial velocities perpendicular to that axis. This
allowed us to recover double-frequency orbits in double bars that correspond
to the x1 and x2 orbital families in a single bar. However, as we pointed out
at the beginning of Section 2, the phase-space of initial conditions for
trajectories in double bars has two more dimensions: the initial relative 
position angle of the bars and the initial radial velocity of the particle.
Below, we extend our exploration of trajectories in double bars,
so that the phase-space is sampled to a similar extent as with the surfaces 
of section in a single bar. This can be achieved by constructing two more 
types of 2-dimensional diagrams: one with a condition of aligned bars replaced
by a given, non-zero initial relative position angle of the bars, and the 
second one by replacing the condition of zero initial radial velocity by a 
fixed angle between the initial velocity vector and the minor axis of two
aligned bars.

\subsection{Starting particles at various angles between the bars}
In order to check whether the double-frequency orbits in 
the Reference Model trap around themselves a significant volume of 
regular orbits continuously, as the bars rotate with respect to each
other, we calculated the ring widths for maps of trajectories that start at 
five more relative orientations of the bars. Each such map consists of points 
on trajectories recorded at that given relative orientation of the bars,
and therefore it can indicate how the trapped trajectories behave as the bars
rotate through each other.

We took as a reference the minor axis of the outer bar and, as before, we 
started particles from that axis, with the initial velocity perpendicular to 
it. We calculated the diagrams of ring widths for the initial angle between the
bars equal to 0.2$\pi$, 0.4$\pi$, 0.5$\pi$, 0.6$\pi$ and 0.8$\pi$. Since the 
equation of motion is invariant with respect to time reversal, which reverses 
both the velocity of the particles and the rotation of the bars, trajectories 
launched when 
the angles between the bars are $\alpha$ and $\pi - \alpha$ are mirror images
of one another, hence their diagrams of ring width are the same. In fact, 
our diagrams for $\alpha = 0.2\pi$ and 0.8$\pi$ look the same, as well as 
the diagrams for 0.4$\pi$ and 0.6$\pi$, 
which makes a good consistency check. Thus, in Fig.3 we present only the 
diagrams for the relative angles of the bars $\alpha=0.5\pi$, 0.4$\pi$ and 
0.2$\pi$.

The system is symmetric with respect to the axes of both bars only when the
bars are aligned, or orthogonal. One may expect that in these two orientations
trajectories starting orthogonally to the bar's axes play a special role, as 
closed periodic orbits possessing this property do in a single bar. Thus the 
diagrams
for bars aligned and orthogonal can be directly compared (left panels of Figs 1
and 3). In both diagrams one can see the two dark arches of similar shape and 
location, corresponding to the major orbital families. In both cases white 
stripes indicating irregular orbits run across them. The grey regions around
the arches, marking regular trajectories trapped around double-frequency 
orbits, have similar extent in both diagrams, which means that a similar 
amount of initial conditions at bars aligned and orthogonal generates trapped 
trajectories. This observation can be quantified by comparing the histograms
of ring widths for both diagrams, presented in Fig.4. Although the potential
with two bars orthogonal has a slightly different Lagrange radius $r_L$, and
$v_{\rm max}$ in (2) takes slightly different values, these differences are
very small and they do not change the appearance of the histograms. The two 
histograms in Fig.4 look very similar, with the one for orthogonal bars
having a slightly larger proportion of trajectories very well trapped (ring 
width about 5 per cent), and less irregular trajectories (ring widths above 
60 per cent).

\begin{figure}
\centering
\rotatebox{-90}{\includegraphics[width=.6\linewidth]{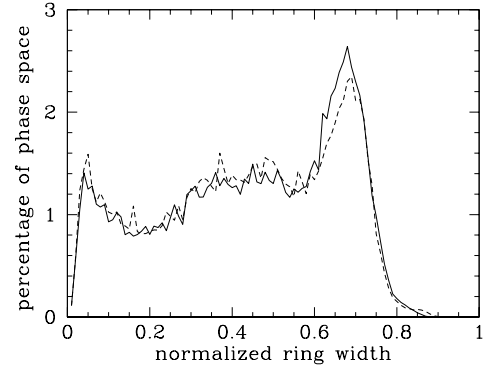}}
\caption[]{Histograms, as in Fig.2, of ring width for maps of trajectories 
in the Reference Model that start when the bars are aligned (solid line), and 
when they are orthogonal (dashed line).}
\label{f4}
\end{figure}

For non-aligned or non-orthogonal bars, there is no symmetry of the system 
around any axis, but orbits dominated by one bar may still map onto loops 
that are symmetric about that bar's axes. This is what we see in the 
diagrams for the angle between the bars equal to 0.4$\pi$ and 0.2$\pi$
(Fig.3, central and right panels).
The outer arch represents orbits corresponding to the x1 orbits in a single
bar, and probably dominated by that bar. The appearance of this arch is 
very similar in all three diagrams in Fig.3. The inner arch marks orbits
corresponding to the x2 orbits in a single bar, but also forming the backbone
of the inner bar in double bars (see MS00). This twofold nature of that
family of orbits is reflected in Fig.3: the right leg of the arch, 
corresponding to the outer orbits, remains largely unchanged in all the three 
diagrams -- these orbits are under the dominating influence of the outer bar.
The left leg of the arch, corresponding to the inner orbits, gets increasingly 
pale as one moves to the diagrams for the angles $\alpha = 0.4\pi$ and 
0.2$\pi$, and it misses its darkest 'spine'. These orbits are dominated by 
the inner bar, and their maps are not symmetric around the outer bar's minor 
axis, for which the diagrams 
were constructed. As the axes of the bars depart from each other, an increasing
radial component in the initial velocity on these orbits is expected, which is
not taken into account in constructing the diagrams in Fig.3. Thus for the 
angles $\alpha = 0.4\pi$ and 0.2$\pi$, in the left 'leg' of the lower arch,
we likely see trajectories trapped around double-frequency orbits that 
themselves are not included in the diagrams. 

It is interesting to point out that the division of the lower arch into
the parts dominated by the inner and the outer bar coincides with the white
stripe crossing that arch. This stripe means discontinuity in the orbital
family, dividing it into two subfamilies: one following the inner, and one the
outer bar. We did not find the source of the instability that this stripe
represents, but it may play a crucial role in separating dynamically the inner 
bar from the rest of the system.
 
\subsection{Trajectories with a radial component in the initial velocity}
Thus far we considered only trajectories starting in the tangential direction, 
i.e. with no initial radial velocity component. In this subsection, we 
sample all other starting directions by constructing and analyzing a set of
ring-width diagrams, each for a given angle $\theta$ between the initial 
velocity vector and the prograde tangential direction. The ring widths are
measured for maps at the moments when the bars are aligned, and the 
trajectories start on the minor axis of the aligned bars.

\begin{figure*}
\centering
\includegraphics[width=0.98\linewidth]{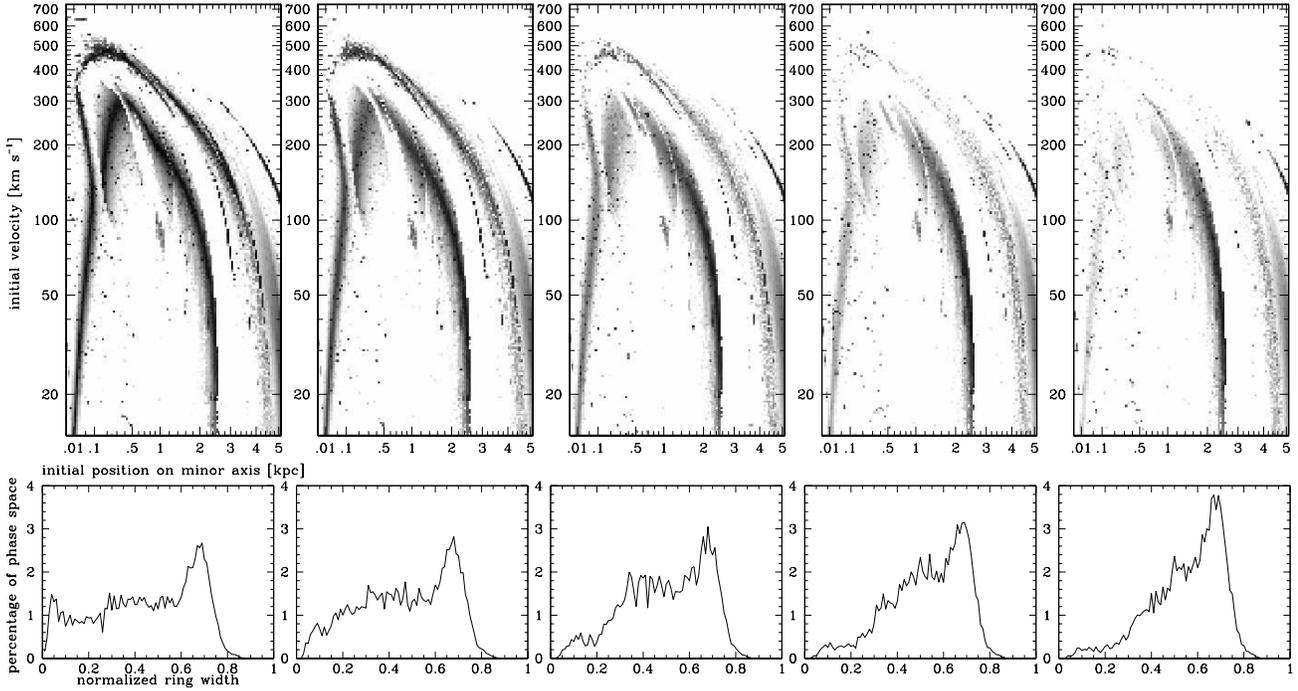}
\vspace{-154mm}
\caption{{\bf Top panels:} Ring-width diagrams in the Reference Model, 
as in the left-hand panel 
of Fig.1, for the maps constructed when the bars are aligned, but for the 
initial velocity vector inclined to the prograde tangential direction at the
angle (left to right diagrams): 0\deg, 3.6\deg, 7.2\deg, 10.8\deg\ and 
14.4\deg. {\bf Bottom panels:} Corresponding histograms of the ring width.}
\end{figure*}

\subsubsection{Stability of double-frequency orbits to small radial velocity
perturbation}
The ring-width diagram from the left panel of Fig.1 already indicates that the
two main families of double-frequency orbits in the Reference Model, 
represented by the two arches there, are stable with respect to small
changes of the starting position on the minor axis of the aligned bars,
and to small changes of the value of the initial prograde tangential 
velocity. In fact this stability is necessary in order to recover the
double-frequency orbits from the ring-width diagrams, like the one in Fig.1.
We still need to check, whether these double-frequency orbits remain stable
when a small radial component in the initial velocity is added.

In order to confirm this, we calculated the ring widths for 
trajectories in the Reference Model as a function of three parameters: the 
starting position on the minor axis of the aligned bars, and the starting 
velocity (its value and direction), around the prograde tangential. With
these three parameters one can describe all orbits that ever pass through the
minor axis of the bars when the bars are aligned, and therefore obtain
information equivalent to what surfaces of section give for a single bar.
The calculated ring-widths constitute a data-cube, which can be displayed
in the form of cuts for the constant direction of the initial velocity, as
in the top panels of Fig.5. There, darkest arches, corresponding to smallest
ring-widths, are found in the diagram for trajectories with no initial radial 
velocity (top-left panel). This means that the major double-frequency orbits
have no radial velocity component when crossing the minor axis of the aligned
bars. As the radial component of the initial velocity (i.e. the angle between 
the initial velocity vector and the prograde tangential direction) increases,
the arches remain in place, although they gradually become brighter. This 
indicates that trajectories, whose initial conditions are the same as those of 
the major double-frequency orbits, except for a radial component in the initial
velocity, remain trapped around those double-frequency orbits, as long as
this radial component is sufficiently small. Trajectories with the initial 
velocity vector inclined to the prograde tangential direction at 
$\theta=3.6$\deg are trapped very well around the double-frequency orbits, but 
the trapping is much less efficient at $\theta=7.2$\deg. Interestingly,
trajectories remain well trapped around the outermost x2 orbits (right
leg of the inner arch) even at $\theta=14.4$\deg (Fig.5, top-right panel).
Efficient trapping of trajectories proves that the major 
double-frequency orbits are stable to small radial velocity perturbations,
and they can serve as the backbone for a doubly barred galaxy.

The lower panels of Fig.5 show the histograms of ring width. The first maximum,
at the ring width of about 5 per cent, indicating trajectories very well
trapped around double-frequency orbits, is notably absent in all histograms,
except for the first one from the left, for which the initial radial velocity
is zero. This again confirms that the major double-frequency orbits cross
the minor axis of the aligned bars perpendicular to it. The characteristic 
quasi-Gaussian component on the right of each histogram, which indicates 
chaotic orbits, increases with the increasing radial component of the initial 
velocity. Overall, however, in all histograms from Fig.5, a considerable 
fraction of phase-space is still populated by regular orbits.

\subsubsection{Exploration of all directions of the initial velocity vector}
In the previous section, we explored trajectories with nearly prograde 
tangential initial velocity. Here we search for double-frequency orbits with a 
significant radial component of the initial velocity, hence we allow any 
initial velocity vector in the galactic plane. We still start the
particle on the minor axis of the aligned bars. We construct a set of 
diagrams, each for a given angle $\theta$ between the initial velocity 
vector and the tangential prograde direction, 
plotting on the diagrams' axes the position on the minor axis of the bars and 
the velocity value. The original diagram from Fig.1 was for $\theta=0$, and we 
constructed diagrams for $\theta=h\pi$, where $h$ runs from -1 to 1 in 
increments of 0.2.

\begin{figure}
\centering
\includegraphics[width=\linewidth]{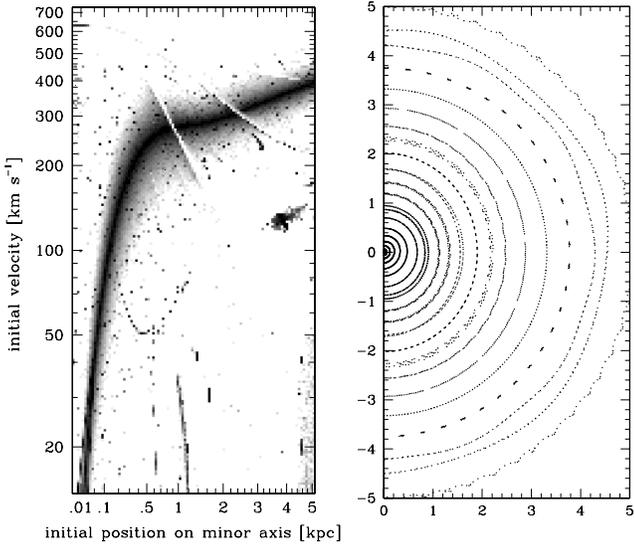}
\vspace{-48mm}
\caption[]{{\bf Left-hand panel:} The ring-width diagram in the Reference 
Model, as in the left-hand panel of Fig.1, for the maps constructed when 
the bars are aligned, but for the initial velocity retrograde.
{\bf Right-hand panel:} Maps of trajectories well trapped around the retrograde
double-frequency orbits (initial conditions taken from the darkest 'spine' of 
the stripe in the left-hand panel), constructed at the
moment when the angle between the bars is $\pi/4$ with the outer bar along
the horizontal axis. Units on axes are in kpc.}
\label{f5}
\end{figure}

In Sect.3.2.1, we showed that as $\theta$ departs from zero, the two arches 
in the top panels of Fig.5 gradually fade into grey, which indicates that there
are regular trajectories with $\theta \neq 0$ still trapped around orbits 
for which $\theta=0$. Here we add that in the area covered by the arches from 
the top-left panel of Fig.5, most of the ring width remains below 50 per cent
in the $\theta=\pm 0.2 \pi$ diagrams, but in the $\theta=\pm 0.4 \pi$ diagrams
ring widths there are mostly above 50 per cent. 
This confirms that our sampling of $\theta$
is dense enough to recover double-frequency orbits that trap around themselves
large regions of phase-space. The $\theta=\pm 0.6 \pi$ diagrams are 
uniformly white -- they contain no trajectories that map onto rings thinner
than half of their radii, and certainly no double-frequency orbits capable of
trapping around themselves any considerable volume of phase-space. The same is 
true for the $\theta=-0.8 \pi$ diagram, while the $\theta=+0.8 \pi$ diagram
shows a very faint signature of orbits trapped around the inner orbits
that manifest themselves in the $\theta=(\pm) \pi$ diagram.

The velocity vector for the $\theta=\pi$ diagram (Fig.6, left panel) is
tangential retrograde, and, predictably, we recover regular trajectories
trapped around the orbital family corresponding to the retrograde x4 family 
in a single bar (Contopoulos \& Papayannopoulos 1980, see also Sellwood \& 
Wilkinson 1993 for a review). Orbits from the darkest spine of the dark
stripe in the diagram map onto loops, which are almost round and are shown 
in the right panel of Fig.6.
We show the loops at the moment when the angle between the two bars is
$\pi/4$, from which one can imply that the slightly elongated loops appear to
be aligned with the inner bar at radii around 0.5 kpc, and with the outer bar
at radii around 3 kpc. However, the highest ellipticity of the loops is only
a few per cent, hence particles placed on these loops cannot recreate the
density distribution as the bars rotate through each other. Since the loops
do not oscillate much, the appearance of the ring-width diagram from the 
left-hand panel of Fig.6 remains the same if we repeat it for other relative 
angles of the bars, similarly to what we did in Sect.3.1 for prograde
trajectories. Note that the diagram from Fig.6 appears smoother than that for 
the tangential prograde velocity from the left panel of Fig.1 -- here it is 
disrupted by
only two narrow white stripes marking irregular orbits, and otherwise
it has a well defined 'spine' with ring width increasing monotonically as one
moves away from it. 

This analysis shows that only parent orbits with the initial velocity 
tangential prograde (counterparts of the x1 and x2 orbits in a single bar) 
and retrograde (counterparts of the x4 family there) trap a significant volume 
of regular trajectories around themselves. The prograde trajectories remain 
trapped for larger range of angles between their initial velocity vector and 
the tangential direction than the retrograde trajectories do. The volume of 
phase-space occupied by the trapped orbits does not change significantly as 
the bars rotate through reach other (Sect.3.1).

\section{Regular motions in 22 new models of double bars}
Here we construct new models of double bars by varying parameters of the 
Reference Model in order to evaluate how the orbital support of double bars 
depends on the parameters of the system.
Using the Reference Model as an example, we showed in the previous section
that double-frequency orbits, which constitute the backbone of a doubly barred
galaxy, originate from the x1 and x2 orbital families in a single bar. These
double-frequency orbits can be found in maps of trajectories that cross the
minor axis of the aligned bars tangentially (i.e. with no radial velocity). 
Such trajectories have only two free initial conditions: the initial position 
on the minor axis of the aligned bars and the initial tangential velocity, and
we calculate here ring widths as a function of these two initial conditions
only.

\begin{table}
\caption{Parameters of the models of double bars}
\begin{tabular}{lcccccccc}
\hline
Model & & \multicolumn{7}{c}{parameters of the}\\
name & & \multicolumn{3}{c}{outer bar} & \multicolumn{4}{c}{inner bar} \\
\hline
   &$\rho_c$&$r_L$&$\frac{a_1}{b_1}$&$Q_M$&$\Omega_2$&$\frac{M_2}{M_1}$&$\frac{a_2}{a_1}$&$\frac{a_2}{b_2}$ \\ 
\hline
01 & 4.8    & 6.0 & 2.5     & 4.5 & 110      & 0.15    & 0.2     & 2.0 \\
02 &        &     &         &     &  80      &         &         &     \\
03 &        &     &         &     &  90      &         &         &     \\
04 &        &     &         &     &  100     &         &         &     \\
05 &        &     &         &     &  120     &         &         &     \\
06 &        &     &         &     &  110     & 0.11    &         &     \\
07 &        &     &         &     &          & 0.13    &         &     \\
08 &        &     &         &     &          & 0.17    &         &     \\
09 &        &     &         &     &          & 0.19    &         &     \\
10 &        &     &         &     &          & 0.15    & 0.16    &     \\
11 &        &     &         &     &          &         & 0.18    &     \\
12 &        &     &         &     &          &         & 0.22    &     \\
13 &        &     &         &     &          &         & 0.24    &     \\
14 &        &     &         &     &          &         & 0.2     & 1.5 \\
15 &        &     &         &     &          &         &         & 2.5 \\
16 &        &     &         & 2.25&          &         &         & 2.0 \\
17 &        &     &         & 9.0 &          &         &         &     \\
18 &        &     & 2.0     & 4.5 &          &         &         &     \\
19 &        &     & 3.0     &     &          &         &         &     \\
20 &        & 5.5 & 2.5     &     &          &         &         &     \\
21 &        & 6.5 &         &     &          &         &         &     \\
22 &        & 7.0 &         &     &          &         &         &     \\
23 & 2.4    & 6.0 &         &     &          &         &         &     \\
\hline
\end{tabular}

\medskip
$\rho_c$ is the central density in \solm kpc$^{-3}$ for the total mass 
distribution in the model, $r_L$ is the radial coordinate in kpc of the Lagrange
point $L_1$ in the outer bar, $\frac{a_1}{b_1}$ is the axial ratio of the
outer bar, $Q_M$ is the quadrupole moment of the outer bar in 
$10^{10}$ \solm kpc$^2$, $\Omega_2$ is the pattern speed of the inner bar in
\kmskpc, $\frac{M_2}{M_1}$ is the mass ratio of the bars, $\frac{a_2}{a_1}$
is the ratio of the major axes of the two bars, $\frac{a_2}{b_2}$ is the axial
ratio of the inner bar. Empty fields in the table indicate values the same as 
in the entry above.
\end{table}

In the simplest approach, each bar has at least 4 free parameters: size, mass,
axial ratio and pattern speed. In the galaxy model there are also parameters 
of other components: the disk and the spheroid. This is a very big parameter 
space. We start its exploration by keeping the size of the outer bar unchanged,
and by changing the other parameters one by one. In effect, we built 23 models,
listed in Table 1, for which ring-width diagrams are presented in Fig.7, 
and histograms of ring width in Fig.8. In each model, volumes of phase-space,
given by (3), that sum up to generate the histogram (see Sect.2.3), are 
normalized to the Lagrangian radius $r_L$ and the velocity $v_{\rm max}$ 
appropriate for that model. Below we analyze the models according to the 
parameters that were varied. Model 01 is our Reference Model.

{\it Models with varying pattern speed of the inner bar, $\Omega_2$}. We varied
$\Omega_2$ between 80 and 120 \kms, with the Reference Model value being 110 
\kms.
Four models belonging to this group (02,03,04,05) are displayed in the left
panels of the first row from the top in Fig.7, and their histograms -- in the
first panel from the left of the top row in Fig.8. From the histograms one 
can see that the slower the inner bar rotates, the higher is the fraction of
phase-space occupied by chaotic orbits. This is reflected in the ring-width 
diagrams in Fig.7, where the disruption of both arches that mark the main 
families of orbits becomes more severe with decreasing pattern speed of the 
inner bar. In models 02 and 03, part of the outer arch is missing around the 
initial velocity about 400 \kms, hence there are no stable orbits to support 
the inner parts of the outer bar. In these same models, the white instability 
strip crossing the inner arch erases most of the stable orbits that were to 
support the inner bar. On the other hand, for pattern speeds higher than that 
in the Reference Model, the outer arch appears to be continuous, and the white
stripe crossing the inner arch is much narrower, not disturbing much the 
orbital family supporting the inner bar. This exercise shows that double bars
with the inner bar rotating faster than that in the Reference Model may be more
dynamically plausible than that model.

{\it Models with varying mass of the inner bar}. We built four models, for 
which the mass of the outer bar is the same as in the Reference Model, while 
the mass ratio of the bars, $\frac{M_2}{M_1}$, varies between 0.11 and 0.19, 
with the Reference Model value 
being 0.15. Four models belonging to this group (06,07,08,09) are displayed in 
the left panels of the second row from the top in Fig.7, and their histograms 
-- in the second panel from the left of the top row in Fig.8. Interestingly,
varying the mass of the inner bar by almost a factor of 2 does not produce
any noticeable difference in the extent of regular and chaotic zones in the 
system. Also ring-width diagrams have similar general appearance for all four
models. However, a detailed look at these diagrams reveals that the upper arch 
becomes discontinuous at around 400 \kms\ for the more massive inner bar. This
again means a disruption of the inner orbits supporting the outer bar. Also 
the right leg of the outer arch is thinner for the more massive inner bar, 
which means that in general, the outer bar traps smaller volume of orbits once 
the inner 
bar becomes more massive. On the other hand, the white stripe crossing the
inner arch becomes wider as the mass of the inner bar decreases, and it erases 
stable orbits supporting the inner bar. Therefore some intermediate mass of the
inner bar, close to value assumed in the Reference Model, is adequate in the 
most regular system. We also note that the white stripe crossing the inner arch
systematically moves to the right, with respect to that arch, as the mass of
the inner bar increases.

\begin{figure*}
\centering
\includegraphics[width=\linewidth]{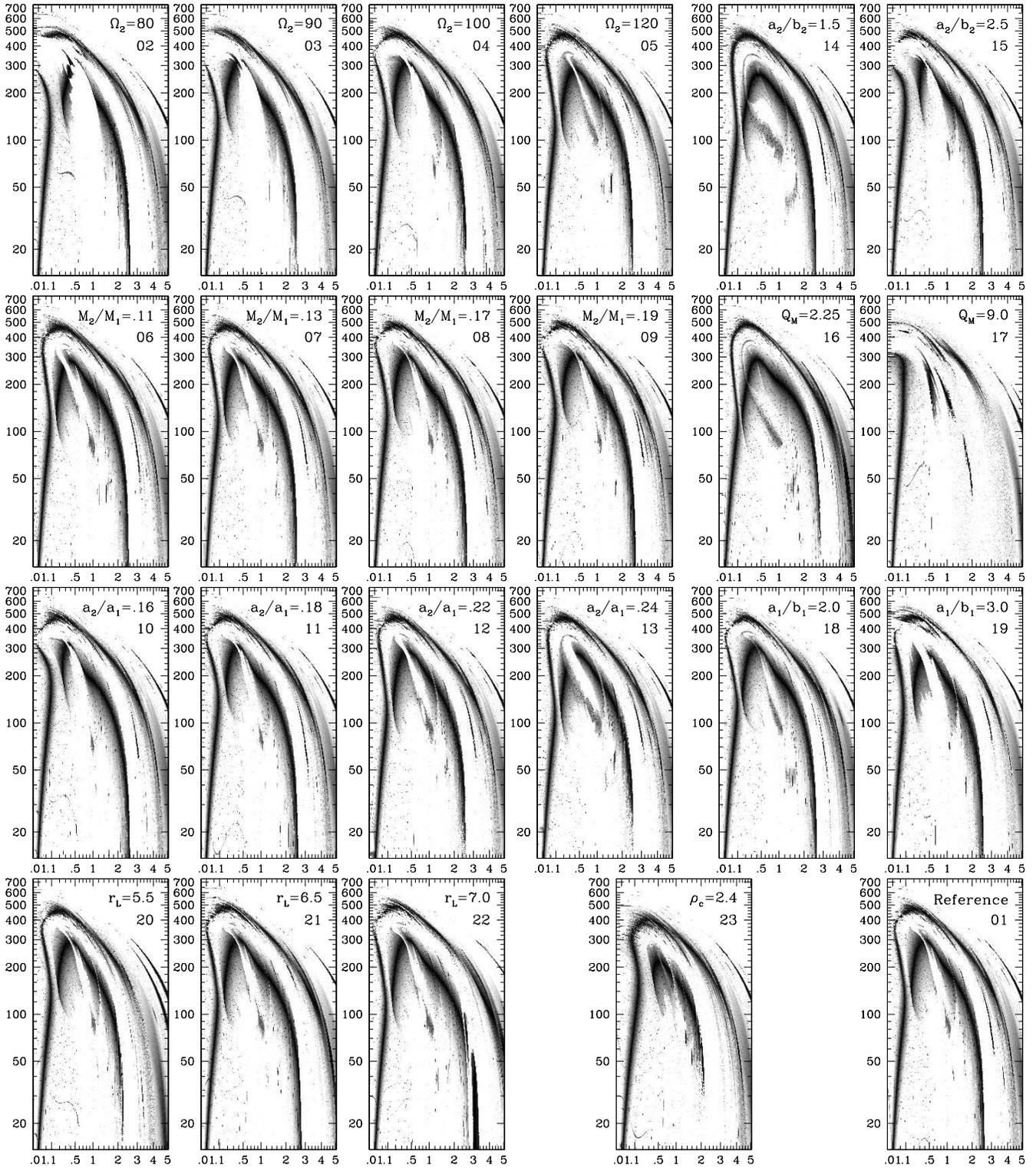}
\vspace{-5cm}
\caption[]{Ring-width diagrams, as in the left-hand panel of Fig.1, for
the Reference Model (bottom-right corner) and 22 further models of 
doubly barred galaxies constructed in this work. The diagrams
are gathered in groups, containing models for which one parameter is
varied. The symbol of the varying parameter (explained in Table 1), together 
with its value, is given in the upper-right corner of each diagram, followed 
by the number of the model below it. As in the left panel of Fig.1, the 
coordinate along the horizontal axis of the diagrams is the initial position 
of the particle on the minor axis of the aligned bars, in kpc, and along the 
vertical axis -- its initial tangential velocity in \kms.}
\label{f6}
\end{figure*}

\begin{figure*}
\centering
\vspace{-7mm}
\includegraphics[width=1.05\linewidth]{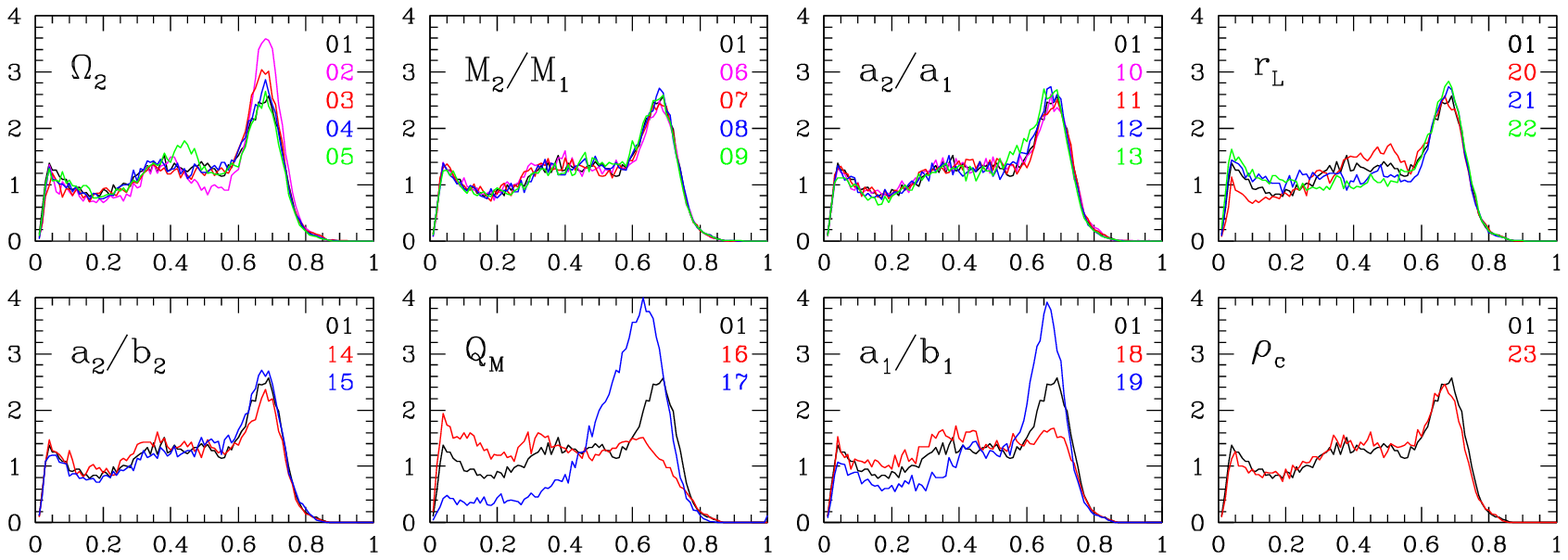}
\vspace{-125mm}
\caption[]{Histograms of ring width for the Reference Model (model 01) and the
22 new models. As in Fig.7, the 
histograms are gathered in groups: each plot displays a set of histograms for
models, in which one parameter was varied. The varied parameter (explained in 
Table 1) is given in the upper-left corner of each plot. The
models are colour-coded with their numbers listed in the plots in the
appropriate colour. As in Fig.2, the coordinate along the horizontal axis of 
the histograms is the normalized ring width, and along the vertical axis -- 
the percentage of phase-space.}
\label{f8}
\end{figure*}

{\it Models with varying size of the inner bar}. We built four models, for 
which the size of the outer bar is the same as in the Reference Model, while 
the size ratio of the bars, $\frac{a_2}{a_1}$, varies between 0.16 and 0.24, 
with the Reference Model value 
being 0.2. Four models belonging to this group (10,11,12,13) are displayed in 
the left panels of the third row from the top in Fig.7, and their histograms 
-- in the third panel from the left of the top row in Fig.8. As in the case
of varying the mass of the inner bar, there is no noticeable difference in the 
extent of regular and chaotic zones among the models. However, the
ring-width diagrams show clear differences. In model 10, in which the inner bar
is the smallest, the outer arch is again disrupted, around the initial 
velocity about 350 \kms. This is likely because we keep the mass of the
inner bar constant between the 4 models considered here, and decreasing the
size of the inner bar means increasing its density, which may lead to 
the disruption of orbits supporting the inner part of the outer bar. Making
the inner bar larger for a given mass brings back the continuity of the outer
arch, but then the stripe crossing the inner arch widens. In model 13, with 
the largest in size inner bar, a large fraction of that arch is being erased, 
weakening the backbone of the inner bar. As a result, some intermediate size 
of the inner bar, close to the value assumed in the Reference 
Model, is adequate in the most regular system, as it also is the case for the 
mass of the inner bar, which we found above. One other remark on this set of
models is that even if the histograms of the models look the same, one 
can differentiate on the plausibility of these models based on their
ring-width diagrams.

{\it Models with varying axial ratio of the inner bar, $\frac{a_2}{b_2}$}. 
We varied $\frac{a_2}{b_2}$ between 1.5 and 2.5, with the Reference Model value
being 2.0. Two models belonging to this group (14,15) are displayed in the 
right
panels of the first row from the top in Fig.7, and their histograms -- in the
first panel from the left of the bottom row in Fig.8. The differences between
the histograms are small, but systematic: the smaller the axial ratio, the 
higher the volume of phase-space well trapped around the double-frequency 
orbits, and the smaller the fraction of the chaotic zones. This finding is 
confirmed by the ring-width diagrams. In the diagram for model 14 with the 
smallest axial ratio of the inner bar, the two arches are continuous, and there
is even no white instability stripe crossing the inner arch. Thus the two main
orbital families are continuously stable, providing good support for the
two bars. This is not unexpected, given that the inner bar is closest to 
being axisymmetric in this model. To the contrary, in the diagram for model 
15, with the biggest axial ratio of the inner bar, the two arches are 
discontinuous. This diagram shows that it may be difficult to construct
models of double bars with a thin inner bar.

{\it Models with varying mass of the outer bar}. We built two models, for which
the size of the outer bar is the same as in the Reference Model, while its 
quadrupole moment, $Q_M$, varies between $2.25 \times 10^{10}$ \solm kpc$^2$ 
and $9.0 \times 10^{10}$ \solm kpc$^2$. Since all other parameters of the bar
remain unchanged, the quadrupole moment is proportional to the bar mass, and
therefore there is a 
factor of 4 difference in the bar mass in our models. Two models belonging to 
this group (16,17) are displayed in the right panels of the second row from 
the top in Fig.7, and their histograms -- in the second panel from the left 
of the bottom row in Fig.8. Both the histograms, and the the ring-width
diagrams differ a lot. Model 17, with the most massive outer bar, shows a very
large fraction of phase-space occupied by chaotic orbits, and the regular 
component of the histogram
is scaled down. Its ring-width diagram shows only scattered fragments of the 
inner arch, and a very disrupted outer arch. This model is certainly
prohibited dynamically, and we imply that it may be difficult to construct
models of double bars with a very massive outer bar. On the other hand,
the histogram for model 16, with the least massive outer bar, shows almost
no chaotic component, and a good fraction of phase-space very well trapped 
around the parent orbits. The arches in its ring-width diagram are continuous.
This model is preferred dynamically, but its outer bar is weaker than in the 
commonly observed double bars. 

{\it Models with varying axial ratio of the outer bar, $\frac{a_1}{b_1}$}.
We varied $\frac{a_1}{b_1}$ between 2.0 and 3.0, with the Reference Model value
being 
2.5. Two models belonging to this group (18,19) are displayed in the right
panels of the third row from the top in Fig.7, and their histograms -- in the
third panel from the left of the bottom row in Fig.8. Unlike the similar change
in the axial ratio of the inner bar, this change has clear consequences for
the fractions of regular and chaotic zones in the system. In model 18, with
the lowest axial ratio of the outer bar, the fraction of phase-space occupied 
by chaotic orbits is 
very small, while in model 19, with this axial ratio highest, the component
of the histogram coming from the chaotic orbits dominates, while the regular
component is scaled down. This is reflected in the ring-width diagrams. In 
model 19, the outer arch is very thin -- a thin outer bar traps around itself
trajectories confined to a very small volume of phase-space, and the inner 
arch is again disrupted by the white
instability stripe. On the other hand, in model 18 both arches are continuous 
and thick enough to indicate good trapping of trajectories by the parent 
orbits. Like in the case of varying mass of the outer bar, we conclude that 
the model with small axial ratio of the outer bar is preferred dynamically, 
but such an outer bar is weaker than in the commonly observed double bars. 

{\it Models with varying radial coordinate of the Lagrange point $L_1$ of 
the outer bar, $r_L$}. We varied $r_L$ between 5.5 and 7 kpc, with the Reference
Model value being 6 kpc. The three models belonging to this group (20,21,22) 
are 
displayed in the left panels of the bottom row in Fig.7, and their histograms 
-- in the right panel of the top row in Fig.8. The differences between
the histograms are small, but systematic: the larger the $r_L$, the larger 
the volume occupied by trajectories well trapped around the double-frequency
orbits, but also the higher the fraction of phase-space occupied by chaotic 
orbits. Thus histograms do not differentiate the models in terms of their 
dynamical plausibility. Moreover, the ring-width
diagrams of the models look very similar. The only difference is seen in 
model 20, which loses orbital support for the outer part of the outer bar, 
since that bar's semi-major axis is larger than its Lagrange radius. 
Trajectories 
in this part of model 20 remain regular, though, which is responsible for
the diminution of low ring widths in its histogram, and a surplus of ring
widths around 0.5. We conclude that the change of $r_L$ has little effect
on the dynamical plausibility of the model.

{\it Models with varying central density of the total mass distribution, 
$\rho_c$}. In addition to varying the parameters of the bars, we varied
$\rho_c$, which is likely to govern the extent of the orbital families. We
constructed model 23 with $\rho_c$ twice smaller than in the Reference Model. 
It is displayed in the bottom row of Fig.7, and its histogram is in the
right panel of the bottom row of Fig.8. Surprisingly, this histogram shows no 
considerable difference from the histogram of the Reference Model. On the other
hand, the appearance of the ring-width diagrams is much different for these two
models. In model 23, the outer arch is thicker, while the extent of the inner
arch is much smaller. This is what one would expect from orbital
structure in a single bar --- for models with smaller central concentration,
i.e. with smaller values of $\rho_c$, the extent of the x2 family is
smaller (e.g. Athanassoula 1992a). Thus one would expect them to have
a more pronounced x1 
contribution and a less pronounced x2, as indeed our calculations
show. Yet, we have explored only two models in the $\rho_c$ sequence
and further work would be useful to understand better how this parameter
affects the structure of double bars.

\section{Discussion}
\subsection{Trends in the orbital response to changing parameters}
In this paper, we explored a limited range of values for all essential
parameters of double bars. We compared the ring-width diagrams and  
histograms of the models. Histograms clearly show a presence of a chaotic 
component, and they can be used to estimate the fraction of phase-space
occupied by chaotic orbits. On the other hand, ring-width diagrams provide 
much more detailed information about which part of phase-space is affected
by changing a given parameter of the model. A histogram provides
information integrated over the ring-width diagram, and models with
diverging structures in ring-width diagrams may produce very similar
histograms. Models 01 and 23, for which the central mass density in
the model was varied, illustrate this case. Although their histograms
do not differ in any appreciable way, their ring-width diagrams are
not at all alike, as they show different extents of the orbital
families, and different volumes of the phase-space trapped around these
families. Smaller extent of the x2 orbits in model 23 and smaller
volume of phase-space trapped around the x1 family in model 01 result
in almost identical integrated distribution in the histograms. Thus, it
is worth stressing that the ring-width diagrams are much more sensitive
indicators than the histograms. Nevertheless, the biggest differences
among our  
models can already be seen in the histograms.

\subsubsection{Size and mass}
Clearly, the fraction of phase-space occupied by regular and 
chaotic orbits changes most, when the parameters of the {\it outer} bar are 
being varied (models 16--17, where $Q_M$ reflects the mass of the outer bar, 
and models 18--19, where the eccentricity of the outer bar is being
varied). In models 16--17, this is partially because there are large
changes of bar mass between the models --- by a factor of 4. On the
other hand, the axial ratio of the outer bar, varying only by a factor
of 1.5 between models 18 and 19, brings spectacular differences
between the ring-width diagrams, which can be witnessed clearly even in
the histograms. The effect of varying these parameters on the onset of
ergodicity is consistent with previous explorations of orbits in a
single bar (e.g. Athanassoula et al.~1983): ``more massive and/or more
eccentric bars create more ergodicity''.

Among the parameters of the inner bar, 
its eccentricity has the same effect as for the outer bar --- it increases
the volume of chaotic motions. The orbital response to varying the 
next two parameters, the ratios of the masses and lengths of the bars,
is not monotonic, as was the case for the parameters considered so far, 
and indicates the existence of optimum values. This is expected, because 
a too massive inner bar will destroy the outer bar, while an inner bar not 
massive enough cannot support itself. For a given mass of the inner bar,
making this bar too small has the same effect as making it too massive for
a given size. Although the
histograms for these parameters do not differ significantly, the
ring-width diagrams indicate the existence of such an optimum (Sect.4),
and this optimum is very close to the parameters of the Reference
Model. Therefore the Reference Model appears to be chosen optimally in
terms of its mass and size.

The response of orbital support in double bars to the variation of the mass 
ratio of the bars was already studied by El-Zant \& Shlosman (2003). They 
explored a two-dimensional surface in the phase-space of all possible 
initial conditions, span by the initial position on the major axis of 
the outer bar and the initial velocity perpendicular to that axis. We
do the same in Sect.4, although our reference is the {\it minor} axis of
the aligned bars, for reasons of continuity with previous work on periodic
orbits. In this way, our study includes orbits self-intersecting on the
major axis, which are common in strong bars (e.g. Athanassoula 1992a),
while self-intersecting on, or near, the minor axis is rare. While for these
initial conditions we calculate ring widths, 
El-Zant \& Shlosman (2003) extract maximal extensions of a trajectory along
the bar and normal to it, which quantifies whether this trajectory supports
the bar. Both approaches indicate that when the inner bar is too massive in
relation to the outer bar, orbits supporting the outer bar are disrupted, and 
that the inner bar cannot sustain its own orbital structure when it is
not massive enough. We find the 
optimal mass ratio to be about 0.15, with 0.11 being probably too small, and
0.19 being rather too large. El-Zant \& Shlosman (2003) find the ratio 0.01
being too small, the ratio 0.04 close to optimal, and the ratio 0.10 being
already too large. The discrepancy of a factor of 
four in the optimal mass ratio arises most likely because El-Zant \& Shlosman 
(2003) refer to a model of a doubly barred galaxy quite different from our
Reference Model: the size ratio of the bars is 0.08 there, compared to our 
0.20, and the ratio of pattern speeds is 8.3, while we adopt 3.1. 
From models 10--13 we see that a smaller 
inner bar should have a smaller optimal mass. We also note that the parameters
of the El-Zant \& Shlosman model correspond to a  much weaker coupling
between the bars than in our Reference Model. 

\subsubsection{Pattern speeds of the bars. Resonant coupling}
Varying the Lagrangian radius of the outer bar (models 20--22) does not 
result in major changes of the orbital structure. It could be claimed that 
this is because we only explored values varying by a factor of
1.27. Nevertheless, this range is substantial since, at least for
galaxies of type no later than SBc, reasonable values of the
corotation radius of the bar can only be found in the range of
1.2$\pm$0.2 of its semi-major axis (Athanassoula 1992b). 

Changing the pattern speed of the inner bar, $\Omega_2$ (models 02--05),
causes significant changes in both histograms and ring-width diagrams, larger
than the variations induced by changing the Lagrangian radius of the outer
bar (models 20--22). The effect of varying $\Omega_2$ is significant, 
since that parameter changes only by a factor of 1.5 between the models. 
The fraction of the histogram occupied by chaotic orbits decreases with
increasing $\Omega_2$, hence our models indicate that in order to minimize 
the contribution from chaotic motions, the corotation of the inner bar has 
to be brought as close as possible to the end of that bar. This result is 
contrary to what happens in a rapidly rotating single bar, where the region 
close to the corotation is associated with chaotic orbits, and therefore 
it increases contribution from chaotic motions. In our models,
however, the inner bar  
extends to at most 60 per cent of its corotation radius 
(see Fig.9), hence it ends well within its ultraharmonic resonance, past 
which chaotic behaviour is expected in a bar.

\begin{figure}
\centering
\includegraphics[width=\linewidth]{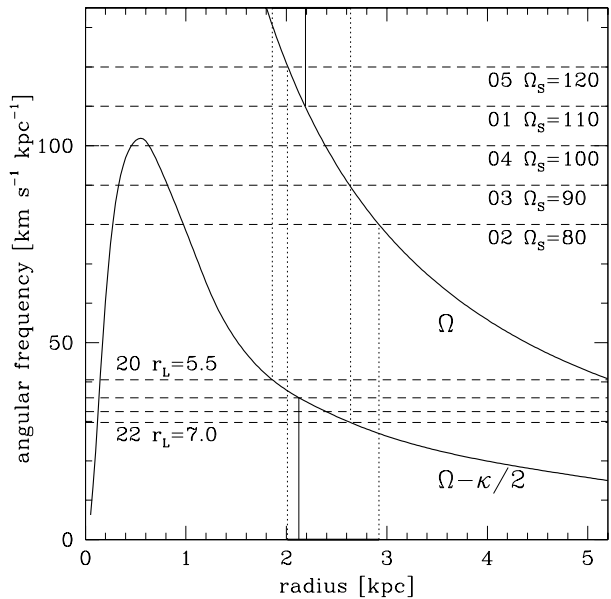}
\caption[]{Thick solid lines mark the angular velocity $\Omega$ (circular 
velocity divided by radius) in azimuthally averaged mass distribution of the 
Reference Model, and $\Omega-\kappa/2$, where $\kappa$ is the free oscillation
frequency. Intersections of these curves with the dashed horizontal lines
of constant pattern speeds of the bars (labeled by the model number, 
followed by the parameter that sets the bar's pattern speed) define the
corotation radii and the positions of the Inner Lindblad Resonances in the
axisymmetric approximation. Two thin solid vertical lines, connecting to the 
upper and lower horizontal axes of the plot, mark the location of the Inner 
Lindblad Resonance of the outer bar (lower one) and the corotation radius of 
the inner bar (upper one) in the Reference Model. Thin dotted vertical lines
mark the range of the Inner Lindblad Resonance of the outer bar and the 
corotation radius of the inner bar in models 02--05 and 20--22. The thick
line segment on the lower horizontal axis
indicates how far the corotation radius 
of the inner bar in the models above departs from the position of the Inner 
Lindblad Resonance of the outer bar in the Reference Model. The thick
line segment on the upper horizontal axis
indicates how far the position of the 
Inner Lindblad Resonance of the outer bar in the models above departs from 
the corotation radius of the inner bar in the Reference Model.}
\label{f9}
\end{figure}

The reason why chaos is reduced in faster rotating inner bars is that for
a corotation radius of the inner bar sufficiently small compared to the 
size of the outer bar, all resonances due to the inner bar fall
in the part where the outer bar is nearly axisymmetric, i.e. the part
where the force from the outer bar is near axisymmetric. This should
decrease the amount of interaction between the two bars and thus also
the amount of chaos. Therefore, a small corotation radius of the inner
bar, i.e. a high $\Omega_2$, should reduce chaos, 
particularly in the inner part of the outer bar, in good agreement  
with what our calculations show. On the other hand, MS00 showed 
that in a doubly barred galaxy, an inner bar extending to its corotation 
has no orbital support. Therefore two opposing factors
shape the ratio of the inner bar's length to its corotation radius,
and in the search for plausible models one should ensure that, in addition to
maximizing the fraction of phase-space occupied by regular orbits, both
bars are supported by loops throughout their extent. 
In Paper III, we will explore loops 
supporting the inner bar in models with various corotation radii, which 
will show whether the optimum should not be shifted in this respect.

The two bars in the Reference Model are roughly in resonant coupling: the 
azimuthally averaged Inner Lindblad Resonance of the outer bar is at 2.13 kpc, 
almost overlapping with the corotation radius of the inner bar at 2.19 kpc.
Tagger et al. (1987) and Sygnet et al. (1988) proposed that resonant
overlap enhances the coupling between nonlinear modes, even at
reasonably low amplitudes. Such a coupling could also 
reduce the extent of chaotic zones in systems with multiple rotating patterns.
In models 02--05 and 20--22, we varied pattern speed of each bar, which 
resulted in departures
from resonant coupling. As shown in Fig.9, the position of the Inner Lindblad 
Resonance of the outer bar varies by factor 1.42 between 1.86 kpc in model 20
and 2.64 kpc in model 22, while the corotation radius of the inner bar varies 
by factor 1.45 between 2.01 kpc in model 05 and 2.92 kpc in model 02. Although
departures from resonant coupling in our models are small, we do not see the
expected minimization of chaos at resonant coupling. To the contrary, 
increasing $\Omega_2$ above the value set by resonant coupling results in
an increase of regular motions in the system. This is in agreement with 
numerical simulations of nearly steady-state double bars (Rautiainen \& Salo 
1999; Shen \& Debattista 2008; see Sect.5.2), which show no preference for
resonant coupling between the bars.

From Fig.9 one can also notice that in the axisymmetric approximation, the
$\Omega-\kappa/2$ curve intersects with the horizontal lines marking pattern 
speeds of the inner bar in models 02--04, which indicates the presence of an
Inner Lindblad Resonance in the inner bar in these models. However, the
ring-width diagrams do not indicate an additional orbital family associated
with this resonance. This shows limitations of the axisymmetric approximation,
already pointed out for a single bar, when the x2 orbits can be absent despite 
the presence of the azimuthally-averaged Inner Lindblad Resonance 
(e.g. Athanassoula 1992a).

\subsection{Comparison with simulations of long-lasting double bars}
Double bars are observed in a large fraction of barred galaxies -- one
third to one fourth according to Erwin \& Sparke (2002) and Laine et al.
(2002), which is consistent with them being being relatively long-lasting.
If double bars last considerably longer than the time between their 
alignments, they can be approximated as oscillating systems.
In this series of papers, we study the orbital support for the oscillating 
potential of double bars. Like the work on orbital structure in single bars, 
our models are not self-consistent, since we assume the potential in
which we calculate the trajectories. A more complete picture of double bar 
dynamics should come from fully self-consistent N-body simulations of 
long-lasting double bars, for which our models can then provide orbital 
structure. 

However, relating the orbital structure in oscillating potentials to 
numerical models is complicated by the fact that double bars which form 
in models including gas are often transient 
features: in the original models of Friedli \& Martinet (1993), the 
decoupled bars survive no more than 7 alignments before they
dissolve. In the simulation of Heller, Shlosman \& Englmaier
(2001) a gaseous ring, there referred to as a nuclear bar, 
tumbles retrograde in the outer
bar's frame for 9 alignments, before settling on a librating
motion. When self-gravity of the gas is included, the ring-bar 
shrinks to the gravitational softening limit within 6 alignments of 
the bars (Englmaier \& Shlosman 2004). All these modeled double bars,
reviewed by Shlosman et al. (2005), are 
transient and comparing their time-dependent parameters with parameters
of our models is not straightforward. Moreover, transient double bars 
cannot account for the large fraction of double
bars observed at the present epoch. 

More recent simulations of Heller, Shlosman \&
Athanassoula (2007a,b) show the longest-lasting double bars to date
for models including a dissipative component. They use
cosmological initial conditions, without any a priori assumptions on
the parameters of the system. The inner bar survives there for at
least 14 alignments of the bars, although it is far from steady-state:
its amplitude decreases roughly three-fold during this time. The weakening
of the inner bar can be caused by mass accumulation in the galaxy centre,
following large radial inflow of gas reported by Heller et al. (2007a).
Pfenniger \& Norman (1990) find limit cycles or strange attractors enhancing 
the nuclear bar in their study of orbits of dissipative particles in double 
bars. 

Long-lived double bars have been first reported in
collisionless (stellar) systems by Rautiainen \& Salo (1999). Their nuclear 
bars form first, survive several gigayears and rotate faster than the outer 
bars. Purely stellar double bars also formed in N-body models by Pfenniger
(2001). More recently,
Debattista \& Shen (2007) and Shen \& Debattista (2008) generated double bars 
in a purely stellar system. In their models, the inner bar forms from the 
bulge that is put in rotation by reversing velocities of particles with 
negative angular momenta. The inner bar lasts for 20 and more alignments,
showing no decay in amplitude, hence it represents a true steady-state of the
system.

\begin{table}
\caption{Parameters of double bars in models and numerical simulations}
\begin{tabular}{lcccc}
\hline
   & Heller & Shen \&    & Reference & models \\
   & et al. & Debattista & Model     & 02-23 \\
\hline
$CR_1/a_1$          & 1.1--1.6  & 1.0--1.4 & 1.0   & 0.9--1.2 \\
$CR_2/a_2$          & 5.0       & 3.25     & 1.83  & 1.5--2.5 \\
$a_2/a_1$           & .09--.14  & .12--.19 & .20   & .16--.24 \\
$\Omega_2/\Omega_1$ & 2.7       & 1.6--2.0 & 3.06  & 2.2--3.3 \\
$CR_1/CR_2$         & 2.3       & 1.5--1.8 & 2.7   & 2.0--3.0 \\
\hline
\end{tabular}

\medskip
$CR_1$ and $CR_2$ are the corotation radii of the outer and the inner
bar, respectively. The meaning of the other symbols is the same as in
Table 1 and in the main text.
\end{table}

In Table 2, we compare the relative sizes of the bars and their rotation rates 
in our models with those extracted from the numerical simulations by Heller et 
al.~(2007a) and Shen \& Debattista (2008). The corotation of the outer bar, 
$CR_1$, in our models is given by its Lagrange radius, $r_L$. We use the 
values from the last, third phase of evolution of the Heller et al.~(2007a) 
model, when the system is more stationary. Bars in the model by Shen \& 
Debattista (2008, Model D) rotate with varying rate. We chose the $CR/a$ 
values in Table 2 for their minimal corotation
radii, since there may be no orbital support for bars at higher radii.

As a single bar can extend almost to
its corotation, MS00 tried to construct a model of double bars, in
which also the inner bar extends to its corotation. Under the resonant
coupling that they assumed (corotation of the inner bar on top of the
Inner Lindblad Resonance of the outer bar), it turned out impossible:
there were no orbits that could support the shape of the inner bar
near its corotation. MS00 concluded that a self-consistent inner bar
must end well inside of its own corotation. Now, the numerical simulations
listed in Table 2 fully confirm this result: the $CR_2/a_2$ ratio
there is even larger than in the Reference Model. This is because
the Reference Model was constructed by MS00 in search for the largest possible
inner bar. Its $a_2/a_1$ ratio is slightly larger than in the
numerical models, and it is situated at the top end of the observed values 
(Erwin \& Sparke 2002, Laine et al.~2002, Erwin 2004). A bar extending to
its corotation radius incorporates higher-order orbital families (3:1, 4:1
etc.) that contribute to its shape. In a bar ending well within its
corotation there is no such contribution. This suggests that the
morphologies of the two bars in the dynamically possible double-barred systems
should differ, while the implications from the observations are
quite opposite: Erwin (2005) points out at many similarities between 
the inner and the outer bar in his images of doubly barred galaxies.

The ratio of pattern speeds (or corotation radii) in our models encompasses
that in the Heller et al.~model, but values in the Shen \& Debattista
model are somewhat lower. This may be a result of arbitrary initial
conditions for the bulge, out of which the inner bar forms in that
model. The ratio of the extent of the outer bar to its corotation
radius consistently takes values around 1 in both simulations and in 
our models. Overall, the agreement between the parameters of our models
and of double bars in numerical simulations indicates that the simulated 
systems are in fact supported by stable double-frequency orbits. 

\section{Conclusions}
We conducted here an extensive survey of trajectories in a potential of
double bars, with completeness similar to that of surfaces of
section in a single bar. Using Model 2 from MS00 as our Reference Model,
we found that only double-frequency orbits
related to the fundamental x1, x2 and x4 orbits in a single bar trap
around themselves the volume of regular orbits large enough to form 
the backbone of the double bar. Similar volumes of phase
space are trapped around backbone orbits in single and double bars, hence
the backbone of double bars appears as robust as in a single bar.

We also constructed 22 further models 
of double bars in order to study how the continuity
of the fundamental x1 and x2 orbital families, and their ability to
trap regular trajectories changes with varying the most essential
parameters of the system. We found that the parameters of our Reference
Model are optimal with those respects: the inner bar has an optimal size for 
its given mass and vice versa. Otherwise, increasing the eccentricity of the 
inner bar increases the fraction of phase-space occupied by irregular
orbits, as is already known for the outer bar. 

The ratio of inner bar's
size to its corotation radius must be such that there are orbits
supporting the inner bar throughout its extent, but also such that
chaotic zones, breaking the continuity of orbital families, are not too
large. Since one condition requires a small ratio, while the other, a
large one, only when these two conditions can be reconciled, the double bar 
is dynamically possible. There is no obvious relation of the conditions 
above to the postulated resonant coupling between the bars. In fact, we
observe reduced chaos in a model that departs from resonant coupling,
which indicates that resonant coupling may not minimize chaos in
double bars.

Recent numerical simulations produced double bars that
last long enough to implicate underlying periodicity in the potential,
thus making it possible to test predictions of the orbital
analysis. The main prediction of MS00, that the inner bar should end
well within its corotation is fully confirmed by these simulations,
which reinforces the predictive power of orbital studies. 
Moreover, in the simulated bars, the corotation radius of
the inner bar can be as far as five times its extent. This calls for
extending the analysis presented here to finding orbital support of
the simulated systems.

{\bf Acknowledgments.} This work was supported by the Polish Committee for 
Scientific Research as a research project 1 P03D 007 26 in the years 
2004--2007 (WM), and by the grant ANR-06-BLAN-0172 in the years 2006--2007
(EA). It was in part carried out within the framework of the 
European Associated Laboratory ``Astrophysics Poland-France".

\end{document}